\documentclass{ws-ijmpa}
\usepackage{amsmath}
\usepackage[super,compress]{cite}
\usepackage{graphicx}
\usepackage{hyperref}
\usepackage{mathrsfs}
\def\e{\mbox{\boldmath$\displaystyle\mathbf{\epsilon}$}}
\def\ep{\mbox{\boldmath$\displaystyle\mathbf{\epsilon}$}}
\def\p{\mbox{\boldmath$\displaystyle\mathbf{p}$}}

\def\q{\mbox{\boldmath$\displaystyle\mathbf{q}$}}
\def\bv{\mbox{\boldmath$\displaystyle\mathbf{\varphi}$}}

\def\0{\mbox{\boldmath$\displaystyle\mathbf{0}$}}
\def\1{\mbox{\boldmath$\displaystyle\mathbf{1}$}}
\def\s{\mbox{\boldmath$\displaystyle\mathbf{\sigma}$}}
\def\J{\mbox{\boldmath$\displaystyle\mathbf{J}$}}

\def\x{\mbox{\boldmath$\displaystyle\mathbf{x}$}}

\def\x{\mbox{\boldmath$\displaystyle\mathbf{x}$}}
\def\y{\mbox{\boldmath$\displaystyle\mathbf{y}$}}
\def\be{\begin{equation}}
\def\ee{\end{equation}}
\def\bea{\begin{eqnarray}}
\def\eea{\end{eqnarray}}

\newcommand{\dual}[1]{\overset{{}^{{}^{\boldsymbol{\neg}}}}{\smash[t]{#1}}}
\newcommand{\gdual}[1]{\overset{\:{}^{{}^{\boldsymbol{\neg}}}}{\smash[t]{#1}}}

\begin{document}
\markboth{Cheng-Yang Lee}
{Self-interacting mass dimension one fields for any spin}

%
\catchline{}{}{}{}{}
%

\title{SELF-INTERACTING MASS DIMENSION ONE FIELDS\\ FOR ANY SPIN
}

\author{CHENG-YANG LEE
}

\address{Institute of Mathematics, Statistics and Scientific Computation,\\
Unicamp, 13083-859 Campinas, S\~{a}o Paulo, Brazil.}
\address{
Department of Physics and Astronomy,\\
Rutherford Building, University of Canterbury \\ 
Private Bag 4800, Christchurch 8140, New Zealand
\\
cylee@ime.unicamp.br}

%

\maketitle

\begin{history}
\received{Day Month Year}
\revised{Day Month Year}
\end{history}

\begin{abstract}
According to Ahluwalia and Grumiller, massive spin-half fields of mass-dimension one can be constructed using the eigenspinors of the charge-conjugation operator (Elko) as expansion coefficients. In this paper, we generalize their result by constructing quantum fields from higher-spin Elko. The kinematics of these fields are thoroughly investigated. Starting with the field operators, their propagators and Hamiltonians are derived. These fields satisfy the higher-spin generalization of the Klein-Gordon but not the Dirac equation. Independent of the spin, they are all of mass-dimension one and are thus endowed with renormalizable self-interactions. These fields violate Lorentz symmetry. The violation can be characterized by a non Lorentz-covariant term that appears in the Elko spin-sums. This term provides a decomposition of the generalized higher-spin Dirac operator in the momentum space thus suggesting a possible connection between the mass-dimension one fields and the Lorentz-invariant fields.

\end{abstract}

\ccode{PACS numbers:}


\section{Introduction}	

The foundation of the Standard Model of particle physics (SM) is closely connected with the Dirac field and its symmetries. The experimental verification of the SM and the elegance of the Dirac equation, have implicitly led us to assume that all massive spin-half fields must satisfy the Dirac equation and are of mass-dimension three-half. This assumption has now been called into question due to the works of Ahluwalia and Grumiller~\cite{Ahluwalia:2004ab,Ahluwalia:2004sz} and it has since been further developed in Refs.~\citen{Ahluwalia:2009rh,Ahluwalia:2008xi}. These works proved by construction, the existence of self-interacting massive spin-half fields of mass-dimension one and that they satisfy the Klein-Gordon but not the Dirac equation. In this paper, we generalize these results by constructing self-interacting mass-dimension one fields for any spin.

The  massive spin-half fields of mass-dimension one are constructed using spin-half Elko as expansion coefficients. Elko is a German acronym for \textit{Eigenspinoren des Ladungskonjugationsoperators}. In English, it means eigenspinor of the charge conjugation operator. The precise mathematical definition of Elko for any spin and the proof of its existence are given in the next section. 

Elko is closely related to the Majorana spinors. The term Elko was only later introduced to avoid possible confusions. A detailed exposition on their similarities and differences can be found in Ref.~\citen{Ahluwalia:2009rh}. Prior to the introduction of Elko, the properties of the higher-spin generalization of Majorana spinors have been investigated by Ahluwalia~\cite{Ahluwalia:1994uy}. While many results obtained by Ahluwalia are in agreement with the results presented in this paper, due to additional insights, the kinematics and the physical interpretation of Elko and the massive spin-half fields now differ from the original ones given in Ref.~\citen{Ahluwalia:1994uy}.

The spin-half Elko have the correct transformation under the $(\frac{1}{2},0)\oplus(0,\frac{1}{2})$ representation of the Lorentz group, but their quantum fields are not Lorentz-covariant~\cite{Ahluwalia:2008xi,Ahluwalia:2009rh}.
This result was later generalized to Elko for any spin.~\cite{Lee:2013cwa}\footnote{We are using the term \textit{spin} loosely. By a spin-$j$ field, it is meant that the spinors transform under the finite dimensional $(j,0)\oplus(0,j)$ representation.}  This immediately raises the question on whether the quantum fields under consideration are well-defined. A partial answer was provided by Ahluwalia and Horvath~\cite{Ahluwalia:2010zn}. Their results suggest that the quantum fields satisfy the symmetry of very special relativity (VSR) proposed by Cohen and Glashow~\cite{Cohen:2006ky}.  

%
%
%

The Lorentz violation for the massive spin-half fields of mass-dimension one can be best characterized by a non Lorentz-covariant term that appear in the Elko spin-sums. Ahluwalia~\cite{Ahluwalia:2013uxa} and Speran\c{c}a~\cite{Speranca:2013hqa} have shown that this term provides an interesting decomposition of $\gamma^{\mu}p_{\mu}$ into a product of two non Lorentz-covariant matrices such that all three are mutually commutative. In this paper, we show that their results can be generalized to any spin thus giving us a decomposition of $\gamma^{\mu_{1}\cdots\mu_{2j}}p_{\mu_{1}}\cdots p_{\mu_{2j}}$.

The spin-half Elko and their fields of mass-dimension one have attracted interests from various areas of research. In cosmology, it was shown by various authors that Elko has the properties to generate inflation as well as acting as a source of dark energy~\cite{Boehmer:2010ma,Boehmer:2010tv,Boehmer:2009aw,Boehmer:2008ah,Boehmer:2008rz,Boehmer:2007ut,
Boehmer:2007dh,Boehmer:2006qq,Chee:2010ju,Wei:2010ad,Shankaranarayanan:2010st,Shankaranarayanan:2009sz,
Gredat:2008qf,S.:2014dja,Pereira:2014wta,Basak:2012sn,daSilva:2014kfa,Basak:2014qea}. In addition, their gravitational interactions have also been explored in the scenarios of black holes and braneworlds~\cite{Jardim:2014xla,daRocha:2014dla,Fabbri:2014foa}. The mathematical properties of Elko have been studied in detail by da Rocha et al.~\cite{daRocha:2005ti, daRocha:2008we,daRocha:2007pz,HoffdaSilva:2009is,daRocha:2009gb,daRocha:2011yr,daRocha:2011xb,
Bernardini:2012sc,daRocha:2013qhu,Ablamowicz:2014rpa,Bonora:2014dfa}. Their particle signatures at the LHC and in cosmology have also been studied.~\cite{Dias:2010aa,Alves:2014kta,Agarwal:2014oaa}
Wunderle and Dick have used Elko to construct
supersymmetric Lagrangians for spin-half fields of mass dimension one~\cite{Wunderle:2010yw}. Fabbri has shown that the massive spin-half field of mass-dimension one does not violate causality ~\cite{Fabbri:2009aj,Fabbri:2009ka,Fabbri:2010va} while Ahluwalia, Lee and Schritt studied its symmetries in~Refs.\citen{Ahluwalia:2008xi,Ahluwalia:2009rh,Lee:2013cwa,Nikitin:2014fga}. 

The paper is organized as follow. In sec.~\ref{Elko_any_spin}, we construct Elko for any spin. This includes all the essential properties of Elko that are needed to construct a quantum field, a task that is carried out in sec.~\ref{field_operators}. From the derived propagators, we conclude that the fields are of mass-dimension one independent of their spin. For the Hamiltonians to be physical, the definitions of the conjugate momenta must be modified. The resulting bosonic and fermionic Hamiltonians are positive-definite but now have positive and vanishing vacuum energies respectively. 
Finally, gathering all the relevant results, we show that $\gamma^{\mu_{1}\cdots\mu_{2j}}p_{\mu_{1}}\cdots p_{\mu_{2j}}$ can be decomposed into a product of two non Lorentz-covariant matrices. This decomposition generalizes the result obtained by Ahluwalia~\cite{Ahluwalia:2013uxa} and Speran\c{c}a~\cite{Speranca:2013hqa} and points toward a possible non-trivial connection between mass-dimension one fields and Lorentz-invariant fields.

\section{Elko for any spin}\label{Elko_any_spin}
We follow the procedure outlined in Refs.~\citen{Ahluwalia:2004ab,Ahluwalia:2004sz,Ahluwalia:2008xi,Ahluwalia:2009rh} to construct Elko for any spin. The first task is to define a suitable charge conjugation operator $\mathcal{C}$ in the $(j,0)\oplus(0,j)$ representation of the Lorentz group. Here, we define this operator as~\footnote{See~\ref{A} for details on the derivation of $\mathcal{C}$.}
\begin{equation}
\mathcal{C}\equiv\left(\begin{matrix}
O & -i\Theta^{-1} \\
-i\Theta & O \end{matrix}\right)\mathcal{K}
\end{equation}
where $\mathcal{K}$ is the complex conjugation operator acting to the right. The matrix $\Theta$ is the Wigner time-reversal operator of dimension $(2j+1)\times(2j+1)$ defined by
\begin{equation}
\Theta\J\Theta^{-1}=-\J^{*}\label{eq:Theta}
\end{equation}
where $\J=(J_{1},J_{2},J_{3})$ are the rotation generators of the $(j,0)$ and $(0,j)$ representations. We adopt the basis where $J_{3}$ is diagonal so that $\J$ are given by~Ref.\cite[eqs.~(2.5.21-2.5.22)]{Weinberg:1995mt}
\begin{equation}
(J_{1}\pm iJ_{2})_{\sigma\bar{\sigma}}
=\delta_{\sigma,\bar{\sigma}\pm1}\sqrt{(j\mp\bar{\sigma})(j\pm\bar{\sigma}+1)},\nonumber
\end{equation}
\begin{equation}
(J_{3})_{\sigma\bar{\sigma}}=\sigma\delta_{\sigma\bar{\sigma}}\label{eq:r3}
\end{equation}
where $\sigma=-j\cdots j$. To prove the existence of $\Theta$, we solve for each of the component of $\J$ and apply the complex conjugation operation. This gives us the following identity
\begin{equation}
\J^{*}_{\bar{\sigma}\sigma}=-(-1)^{\bar{\sigma}-\sigma}\J_{-\bar{\sigma},-\sigma}.
\end{equation}
Rewriting this equation in matrix form, it becomes equivalent to eq.~(\ref{eq:Theta}). The general solution for $\Theta$ (up to a sign) is then
\begin{equation}
\Theta_{\bar{\sigma}\sigma}=(-1)^{-j-\bar{\sigma}}\delta_{-\bar{\sigma},\sigma}\label{eq:Theta_soln}
\end{equation}
which guarantees the existence of the Wigner time-reversal operator for any spin. We define $\Theta$ such that when $j=\frac{1}{2}$, $\Theta=-i\sigma_{2}$ where $\sigma_{2}$ is the second component of the Pauli matrix.

We are now ready to construct Elko for any spin. To avoid confusion, objects that transform under the $(0,j)$ or $(j,0)$ representation are referred to as `functions'. The term spinors is strictly reserved for objects that transform under the $(\frac{1}{2},0)$ or $(0,\frac{1}{2})$ representation. Consequently, the German acronym for higher-spin Elko is
\textit{Eigenfunktionen des Ladungskonjugationsoperators}. We start the construction by adopting the standard polarization basis, introducing a set of functions $\phi(\0,\sigma)$ of the $(0,j)$ representation as eigenfunctions of $J_{3}$. 
\begin{equation}
J_{3}\phi(\0,\sigma)=\sigma\phi(\0,\sigma)\label{eq:phi0}
\end{equation}
where $\sigma=-j,\cdots, j$ spanning the eigenvalues of $J_{3}$. In the component form, we take the solutions of $\phi(\0,\sigma)$ (up to a constant) to be
\begin{equation}
\phi_{\ell}(\0,\sigma)=\sqrt{\frac{m}{2}}\delta_{\ell\sigma}
\end{equation}
with $m$ being the mass of the particle. The functions of arbitrary momentum $\phi(\p,\sigma)$ are obtained by 
\begin{eqnarray}
\phi(\p,\sigma)&=&\exp\left[i\mathbf{K}^{(0,j)}\cdot\bv\right]\nonumber\\
&=&\exp(-\J\cdot\bv)\phi(\0,\sigma)\label{eq:phi}
\end{eqnarray}
where $\mathbf{K}^{(0,j)}=i\J$ in the $(0,j)$ representation. The rapidity parameter $\bv$ is given by
\begin{equation}
\cosh\varphi=\frac{\sqrt{|\p|^{2}+m^{2}}}{m},\hspace{0.5cm}
\sinh\varphi=\frac{|\p|}{m},\hspace{0.5cm}
\bv=\varphi\hat{\p}.
\end{equation}
It follows from the existence of $\Theta$ and eq.~(\ref{eq:Theta}) that the set of complex conjugated functions $[\vartheta\Theta\phi^{*}(\0,\sigma)]$ where $\vartheta$ is a phase to be determined transform under the $(j,0)$ representation
\begin{eqnarray}
[\vartheta\Theta\phi^{*}(\p,\sigma)]&=&\exp\left[i\mathbf{K}^{(j,0)}\cdot\bv\right][\vartheta\Theta\phi^{*}(\0,\sigma)]\nonumber\\
&=&\exp\left[\J\cdot\bv\right][\vartheta\Theta\phi^{*}(\0,\sigma)]\label{eq:Theta_phi}
\end{eqnarray}
where $\mathbf{K}^{(j,0)}=-i\J$.
Acting $J_{3}$ on $[\vartheta\Theta\phi^{*}(\0,\sigma)]$, we see that they have opposite spin-projection with respect to $\phi(\0,\sigma)$
\begin{equation}
J_{3}[\vartheta\Theta\phi^{*}(\0,\sigma)]=-\sigma [\vartheta\Theta\phi^{*}(\0,\sigma)].\label{eq:opposite_projection}
\end{equation}
Combining the functions given by eqs.~(\ref{eq:phi}) and (\ref{eq:Theta_phi}), we obtain a set of $2\times(2j+1)$-component functions of the $(j,0)\oplus(0,j)$ representation of the form
\begin{equation}
\chi(\0,\sigma)=\left(\begin{matrix}
\vartheta\Theta\phi^{*}(\0,\sigma)\\
\phi(\0,\sigma)\end{matrix}\right).
\end{equation}
The transformation under boost is given by
\begin{eqnarray}
\chi(\p,\sigma)&=&\kappa(\p)\chi(\0,\sigma)\nonumber\\
&=&\left(\begin{matrix}
\exp(\J\cdot\bv) & O \\
O & \exp(-\J\cdot\bv) \end{matrix}\right)\chi(\0,\sigma)\label{eq:boost}.
\end{eqnarray}
These functions become eigenfunctions of the charge conjugation operator and hence Elko for any spin when $\vartheta=\pm i$
\begin{equation}
\mathcal{C}\chi(\0,\sigma)\vert_{\vartheta=\pm i}=\mp(-1)^{2j}\chi(\0,\sigma)\vert_{\vartheta=\pm i}.\label{eq:cchi}
\end{equation}
This relation holds for arbitrary momentum since $\mathcal{C}$ commutes with $\kappa(\p)$ defined in eq.~(\ref{eq:boost}). We note, when $j$ takes half-integral or integral values, the eigenvalues of  Elko are even or odd with respect to the signs of the phase $\vartheta$. 
It is important to note that this is not in contradiction with the result obtained in Ref.~\cite[eq.~(8)]{Ahluwalia:1994uy}. The spin-one bosonic charge-conjugation matrix given there differs from $\mathcal{C}^{(j=1)}$ up to a multiplication of a constant matrix. Once this differences is accounted for, they become identical.

In order to construct quantum fields, we follow the general structure that a field operator must have two sets of expansion coefficients associated with its particle and anti-particle. Therefore, we introduce the self-conjugate ($\vartheta=i$)  and anti-self-conjugate ($\vartheta=-i$) functions
\begin{subequations}
\begin{equation}
\xi(\0,\sigma)=\chi(\0,\sigma)\vert_{\vartheta=i,\phi(\mathbf{0},\sigma)\rightarrow\rho(\mathbf{0},\sigma)}\label{eq:xi}
\end{equation}
\begin{equation}
\zeta(\0,\sigma)=\alpha(\sigma)\chi(\0,-\sigma)\vert_{\vartheta=-i,\phi(\mathbf{0},\sigma)\rightarrow\varrho(\mathbf{0},\sigma)}\label{eq:zeta}
\end{equation}
\end{subequations}
with
\begin{equation}
\rho_{\ell}(\0,\sigma)=\sqrt{\frac{m}{2}}c\delta_{\ell\sigma},\hspace{0.5cm}
\varrho_{\ell}(\0,\sigma)=\sqrt{\frac{m}{2}}d\delta_{\ell\sigma}
\end{equation}
where $c$ and $d$ are proportionality constants to be determined and $\alpha(\sigma)$ is a phase. The reason behind the introduction of $\rho(\0,\sigma)$ and $\varrho(\0,\sigma)$ and the factor of $\frac{1}{\sqrt{2}}$ will become apparent in sec.~\ref{field_operators} when we study the locality structure of the fields and compute the Hamiltonians. 

The spinor labels and the phase $\alpha(\sigma)$ are non-trivial. For the spin-half fields, their solutions are carefully chosen by the demand of locality when the particles are taken to be Majorana fermions~\cite{Ahluwalia:2008xi,Ahluwalia:2009rh}. As for quantum fields of any spin, consistency demands the general solutions for Elko to coincide with the spin-half case when $j=\frac{1}{2}$. However, it is beyond the scope of this paper to analyse the general locality structure and determine $\alpha(\sigma)$ when the particles are Majorana. Instead, we focus our attention on the quantum fields with distinct particles and anti-particles where the locality structure are insensitive to the choice of $\alpha(\sigma)$.

It is important to note that the original spin-half Elko were constructed in the helicity basis where the two component Weyl spinors of the $(0,\frac{1}{2})$ representation are eigenspinors of $\s\cdot\hat{\p}$. It was later discovered that in the polarization basis, in addition to the labels and phases, the rest spinors must also contain information of a preferred plane under boost for the fields to be local~\cite[eqs.~(17-18)]{Ahluwalia:2009rh}.~\footnote{See~\ref{B} for details.} We will have more to say about this issue in sec.~\ref{spin_sums}. But so far, for the sake of simplifying the calculation, it is not necessary to introduce the required phases in the polarization basis. Later, we will obtain Elko in helicity basis by performing a unitary transformation.

\subsection{Dual coefficients}

In the $(j,0)\oplus(0,j)$ representation, the dual coefficients of an arbitrary function $\psi(\p)$ is usually defined as
\begin{equation}
\overline{\psi}(\p)\equiv\psi^{\dag}(\p)\Gamma\label{eq:dirac_dual}.
\end{equation}
By the demand of Lorentz-invariant norm for the functions, the matrix $\Gamma$ is constrained by the following conditions
\begin{equation}
[\Gamma,\J]=\{\Gamma,\mathbf{K}\}=O
\end{equation}
where we have used the fact that $\J$ and $\mathbf{K}$ are Hermitian and anti-Hermitian respectively. Solving the equations and setting the appropriate proportionality constants to unity, we obtain 
\begin{equation}
\Gamma=\left(\begin{matrix}
O & I \\
I & O \end{matrix}\right)\label{eq:gamma}
\end{equation}
where the $I$ represents the $(2j+1)\times(2j+1)$ identity matrix. Under this dual, the inner-product between the functions as well as their quantum fields are Lorentz-invariant thus allowing us to construct Lorentz-invariant Lagrangian densities.

%

However, Lorentz invariance of the dual alone does not imply that it is universally applicable. In fact, eq.~(\ref{eq:dirac_dual}) is not the correct dual for Elko. Explicit computations show that the following inner-products identically vanish
\begin{equation}
\overline{\xi}(\0,\sigma)\xi(\0,\sigma)=\overline{\zeta}(\0,\sigma)\zeta(\0,\sigma)=0.\label{eq:vanishing_norm}
\end{equation}
For the purpose of constructing Lagrangian densities, apart from the Lorentz invariance of the inner-product, the norms of the coefficients must also be non-vanishing and orthonormal. Equation~(\ref{eq:vanishing_norm}) shows that eq.~(\ref{eq:dirac_dual}) is inadequate for Elko and must therefore be replaced. The solution to this problem was first found by Ahluwalia in Ref.~\citen{Ahluwalia:1994uy} and further developed by Ahluwalia and Grumiller~\cite{Ahluwalia:2004ab,Ahluwalia:2004sz}. Here, we show that their result can be trivially generalized to higher-spin. The solution is based on the observation that not all the inner-products under eq.~(\ref{eq:dirac_dual}) are vanishing. The general inner-products are given by
\begin{subequations}
\begin{equation}
\overline{\xi}(\0,\sigma)\xi(\0,\sigma')=\frac{im(-1)^{-j-\sigma}}{2}\left[c^{*2}-(-1)^{-2\sigma}c^{2}\right]\delta_{\sigma,-\sigma'},\label{eq:xi_dual}
\end{equation}
\begin{equation}
\overline{\zeta}(\0,\sigma)\zeta(\0,\sigma')=\frac{-im(-1)^{-j-\sigma}}{2}
\left[d^{*2}-(-1)^{-2\sigma}d^{2}\right]\delta_{\sigma,-\sigma'}. \label{eq:zeta_dual}
\end{equation}
\end{subequations}
In light of this observation, we define the dual coefficients for Elko as
\begin{subequations}
\begin{equation}
\gdual{\xi}(\0,\sigma)\equiv\beta(\sigma)\xi^{\dag}(\0,-\sigma)\Gamma
\end{equation}
\begin{equation}
\gdual{\zeta}(\0,\sigma)\equiv\beta(\sigma)\zeta^{\dag}(\0,-\sigma)\Gamma
\end{equation}
\end{subequations}
where $\beta(\sigma)$ is a phase to be determined. From eqs.~(\ref{eq:xi_dual}) and (\ref{eq:zeta_dual}), we find that
\begin{equation}
\overline{\xi}(\0,\sigma)\xi(\0,-\sigma)=-\overline{\xi}(\0,\sigma\pm 1)\xi(\0,-(\sigma\pm 1))\label{eq:Elko_dual}
\end{equation}
and likewise for the anti-self-conjugate functions. For reasons to become apparent in sec.~\ref{kinematics}, we demand all the self conjugate functions to have the same norm and likewise for their anti-self-conjugate counterpart. This is achieved by demanding $\beta(\sigma)$ to satisfy the recursive relation
\begin{equation}
\beta(\sigma)=-\beta(\sigma\pm1).\label{eq:recursive_beta}
\end{equation}
Taking $\sigma=j$ as a point of reference, we obtain
\begin{equation}
\beta(\sigma)=(-1)^{j-\sigma}\beta(j)\label{eq:recursive_beta2}
\end{equation}
where $\sigma=-j,\cdots,+j$. 

Computing the Elko norms using eqs.~(\ref{eq:Elko_dual}) and (\ref{eq:recursive_beta2}), we find
\begin{subequations}
\begin{equation}
\gdual{\xi}(\0,\sigma)\xi(\0,\sigma')=
\begin{cases}
im\beta(j)(c^{*2}+c^{2})\delta_{\sigma\sigma'}/2 & j=\frac{1}{2},\frac{3}{2},\cdots \\
im\beta(j)(c^{*2}-c^{2})\delta_{\sigma\sigma'}/2 & j=1,2,\cdots \label{eq:xi_norms}
\end{cases}
\end{equation}
\begin{equation}
\gdual{\zeta}(\0,\sigma)\zeta(\0,\sigma')=
\begin{cases}
-im\beta(j)(d^{*2}+d^{2})\delta_{\sigma\sigma'}/2 & j=\frac{1}{2},\frac{3}{2},\cdots \\
-im\beta(j)(d^{*2}-d^{2})\delta_{\sigma\sigma'}/2 & j=1,2,\cdots. \label{eq:zeta_norms}
\end{cases}
\end{equation}
\end{subequations}
We want the Elko norms to be real, so without the loss of generality, we restrict the square of the proportionality constants $c^{2}$ and $d^{2}$ to be either purely real or purely imaginary. For the norms to be non-vanishing, the proportionality constants must then satisfy
\begin{equation}
c^{*2}=-(-1)^{2j}c^{2},\hspace{0.5cm}
d^{*2}=-(-1)^{2j}d^{2}.\label{eq:cd_constants}
\end{equation}
Substituting eq.~(\ref{eq:cd_constants}) into eqs.~(\ref{eq:xi_norms}) and (\ref{eq:zeta_norms}), the Elko norms simplify to
\begin{subequations}
\begin{equation}
\gdual{\xi}(\0,\sigma)\xi(\0,\sigma')=im\beta(j)c^{*2}\delta_{\sigma\sigma'},\label{eq:xi_norm_final}
\end{equation}
\begin{equation}
\gdual{\zeta}(\0,\sigma)\zeta(\0,\sigma')=-im\beta(j)d^{*2}\delta_{\sigma\sigma'}.\label{eq:zeta_norm_final}
\end{equation}
\end{subequations}
The phase $\beta(j)$ and the proportionality constants will be determined in sec.~\ref{field_operators} when we study the properties of the quantum fields.

\subsection{Spin-sums and preferred plane}\label{spin_sums}
In the rest frame, using the functions defined by eqs.(\ref{eq:xi}) and (\ref{eq:zeta}), the spin-sums of interests to us are
\begin{subequations}
\begin{equation}
N(\0)=\sum_{\sigma}\xi(\0,\sigma)\gdual{\xi}(\0,\sigma),\label{eq:xi_spin_sum}
\end{equation}
\begin{equation}
M(\0)=\sum_{\sigma}\zeta(\0,\sigma)\gdual{\zeta}(\0,\sigma).\label{eq:zeta_spin_sum}
\end{equation}
\end{subequations}
These spin-sums are important because they determine the propagator and locality structure of the associated quantum fields. We first compute them in the polarization and then the helicity basis at rest and at arbitrary momentum respectively. 

\subsubsection*{Polarization basis}
As we have mentioned at the end of sec.~\ref{Elko_any_spin}, the spin-half Elko in the polarization basis must have non-trivial phases to ensure the quantum fields are local. At this stage, we do not know what these specific phases are for Elko of any spin. But fortunately, this is not necessary. By computing $N(\0)$ and $M(\0)$, we can later obtain the correct functions and spin-sums of arbitrary momentum in the helicity basis by performing a unitary transformation. The required phases for Elko in the polarization basis can then be obtained as a limit of their counterpart in the helicity basis.

Computing the spin-sums using eqs.~(\ref{eq:xi}) and (\ref{eq:zeta}), we get
\begin{subequations}
\begin{equation}
N(\0)=\frac{1}{2}\left[im\beta(j)c^{*2}\left(\begin{matrix}
I & O \\
O & I \end{matrix}\right)
+m\beta(j)|c|^{2}(-1)^{2j}\left(\begin{matrix}
O & \Theta\\
\Theta & O \end{matrix}\right)\right],\label{eq:N0}
\end{equation}
\begin{equation}
M(\0)=\frac{1}{2}\left[-im\beta(j)d^{*2}\left(\begin{matrix}
I & O \\
O & I \end{matrix}\right)
+m\beta(j)|d|^{2}(-1)^{2j}\left(\begin{matrix}
O & \Theta\\
\Theta & O \end{matrix}\right)\right].
\end{equation}
\end{subequations}
To understand why non-trivial phases are required in the polarization basis, it is sufficient to study the behaviour of $N(\p)$
\begin{eqnarray}
N(\p)&=&\sum_{\sigma}\xi(\p,\sigma)\gdual{\xi}(\p,\sigma)\nonumber\\
&=&\kappa(\p)N(\0)\kappa^{-1}(\p)
\end{eqnarray}
where we have used the general identity $\Gamma\kappa^{\dag}(\p)\Gamma=\kappa^{-1}(\p)$. Expanding $N(\0)$ using eq.(\ref{eq:boost}) and eq.~(\ref{eq:N0}) we obtain
\begin{equation}
N(\p)=\frac{1}{2}\left[im\beta(j)c^{*2}\left(\begin{matrix}
I & O \\
O & I \end{matrix}\right)+m\beta(j)|c|^{2}(-1)^{2j}
\left(\begin{matrix}
O & e^{\mathbf{J}\cdot\boldsymbol{\varphi}}\Theta e^{\mathbf{J}\cdot\boldsymbol{\varphi}}\\
e^{-\mathbf{J}\cdot\boldsymbol{\varphi}}\Theta e^{-\mathbf{J}\cdot\boldsymbol{\varphi}} & O \end{matrix}\right)\right].
\end{equation}
For arbitrary momentum $\p$, the sub-matrices $e^{\pm\mathbf{J}\cdot\boldsymbol{\varphi}}\Theta e^{\pm\mathbf{J}\cdot\boldsymbol{\varphi}}$ are not Lorentz covariant. In fact, due to its non-covariance, the resulting quantum fields would be non-local. These observations are in agreement with the previous results in the literature that the massive spin-half fields violate Lorentz symmetry~\cite{Ahluwalia:2008xi,Ahluwalia:2009rh}. In spite of these difficulties, the work of Ahluwalia and Horvath suggests the fields may satisfy VSR symmetry~\cite{Ahluwalia:2010zn}.
Instead of attempting to determine the underlying symmetry group, we choose to focus on higher-spin fields and their implications on particle physics and the general structure of quantum field theory.

Returning to the problem at hand, we note that due to eq.~(\ref{eq:Theta}), the spin-sum is in fact invariant and covariant if it is boosted along the 1, 3-axis and 2-axis respectively
\begin{equation}
N(p^{1})=N(p^{3})=N(\0),
\end{equation}
\begin{equation}
N(p^{2})=\frac{1}{2}\left[im\beta(j)c^{*2}\left(\begin{matrix}
I & O \\
O & I \end{matrix}\right)
+m\beta(j)|c|^{2}
\left(\begin{matrix}
O & e^{2J_{2}\varphi}\\
e^{-2J_{2}\varphi} & O \end{matrix}\right)\right]\label{eq:1_3_axis_spin_sum}
\end{equation}
where the sub-matrices $e^{\pm 2J_{2}\varphi}$ are Lorentz-covariant~\cite[app.~A]{Weinberg:1964cn}.
According to eq.~(\ref{eq:1_3_axis_spin_sum}), when the spin-projection of the particle coincides with the direction of the boost (along the 3-axis), the spin-sum remains invariant since the functions of the $(0,j)$ representation are eigenfunctions of $J_{3}$. Therefore, in the helicity basis where the spin-projections are aligned along the direction of motion, one would expect the spin-sum to be invariant under a general boost. 




\subsubsection*{Helicity basis}

Here we determine the spin-sums in the helicity basis by applying a similarity transformation on their counterpart in the polarization basis. But to achieve this, we first have to determine the relationships between Elko in both basis. 

In the helicity basis, the functions of the $(0,j)$ representation are defined as
\begin{equation}
\J\cdot\hat{\p}\phi(\ep,\sigma)=\sigma\phi(\ep,\sigma)\label{eq:Jphat_eigenvectors}
\end{equation}
where
\begin{equation}
\hat{\p}=(\cos\phi\sin\theta,\sin\phi\sin\theta,\cos\theta)
\end{equation}
and $\ep=\p\vert_{|\mathbf{p}|\rightarrow0}$ instead of $\p=\0$
so the spin-projections are aligned along $\hat{\p}$. Since $\J\cdot\hat{\p}$ is Hermitian and has the same eigenvalue spectrum as $J_{3}$, it follows that there exists a unitary matrix $S(\theta,\phi)$ (whose precise form is of no interest to us) such that
\begin{equation}
SJ_{3}S^{\dag}=\J\cdot\hat{\p}.\label{eq:J_3_diagonalise}
\end{equation}
Therefore, instead of solving eq.~(\ref{eq:Jphat_eigenvectors}) for the eigenfunctions, we obtain $\phi(\ep,\sigma)$ using eq.~(\ref{eq:J_3_diagonalise}) by performing the following unitary transformation
\begin{equation}
\phi(\ep,\sigma)=S\phi(\0,\sigma).\label{eq:0_j_spinors_helicity}
\end{equation}
Following the same computation that gave us eq.~(\ref{eq:Theta_phi}), we see that 
$\vartheta\Theta\phi^{*}(\ep,\sigma)$
transforms under the $(j,0)$ representation and has opposite helicity with respect to $\phi(\ep,\sigma)$
\begin{equation}
\J\cdot\hat{\p}\left[\vartheta \Theta\phi^{*}(\ep,\sigma)\right]=-\sigma\left[\vartheta \Theta\phi^{*}(\ep,\sigma)\right].\label{eq:j_0_spinors_helicity}
\end{equation}
From eq.~(\ref{eq:J_3_diagonalise}) we also obtain
\begin{equation}
J_{3}\left[S^{-1}\Theta\phi^{*}(\0,\sigma)\right]=-\sigma\left[S^{-1}\Theta\phi^{*}(\0,\sigma)\right]
\end{equation}
thus allowing us to make the following identification 
\begin{equation}
\Theta\phi^{*}(\ep,\sigma)=S\Theta\phi^{*}(\0,\sigma).
\end{equation}
Strictly speaking, this identification holds up to a constant. But for $\chi(\ep,\sigma)$ to be an eigenspinor of the charge conjugation operator, we must take the constant to be unity. Combining the functions given by eqs.~(\ref{eq:0_j_spinors_helicity}) and (\ref{eq:j_0_spinors_helicity}), we obtain
\begin{equation}
\chi(\ep,\sigma)=\left(\begin{matrix}
\vartheta \Theta\phi^{*}(\ep,\sigma)\\
\phi(\ep,\sigma)
\end{matrix}\right).
\end{equation}
Setting $\vartheta=\pm i$ and choosing the relevant phases, we obtain Elko in the helicity basis. Since the definitions of Elko were already given by eqs.~(\ref{eq:xi}) and (\ref{eq:zeta}) and remain unchanged, we refrain from repeating ourselves. Elko in the helicity and polarization basis are therefore related by the following unitary transformation
\begin{equation}
\xi(\ep,\sigma)=\mathcal{S}\xi(\0,\sigma),\hspace{0.5cm}
\zeta(\ep,\sigma)=\mathcal{S}\zeta(\0,\sigma)
\end{equation}
where
\begin{equation}
\mathcal{S}=\left(\begin{matrix}
S & O \\
O & S \end{matrix}\right).
\end{equation}
The self-conjugate spin-sum in the helicity basis is then given by
\begin{eqnarray}
N(\ep)&=&\mathcal{S}N(\0)\mathcal{S}^{\dag}\nonumber\\
&=&\frac{1}{2}\left[im\beta(j)c^{*2}\left(\begin{matrix}
I & O \\
O & I \end{matrix}\right)
+m\beta(j)|c|^{2}(-1)^{2j}
\left(\begin{matrix}
O & S\Theta S^{\dag} \\
S\Theta S^{\dag} & O \end{matrix}\right)\right].
\end{eqnarray}

Now we need to boost $N(\e)$ to arbitrary momentum and determine the precise form of the sub-matrix $S\Theta S^{\dag}$. The former task can be accomplished as follow. Noting that the boost along the 3-axis is
\begin{equation}
\kappa(p^{3})=\left(\begin{matrix}
e^{J_{3}\varphi} & O \\
O & e^{-J_{3}\varphi} \end{matrix}\right),
\end{equation}
using eq.~(\ref{eq:J_3_diagonalise}), the general boost $\kappa(\p)$ can then be obtained by a similarity transformation
\begin{equation}
\kappa(\p)=\mathcal{S}\kappa(p^{3})\mathcal{S}^{\dag}
\end{equation}
where $|\p|=p^{3}$. But since $N(p^{3})=N(\0)$, a straightforward calculation shows that the spin-sum in the helicity basis is boost-invariant
\begin{equation}
N_{h}(\p)=N(\ep).
\end{equation}
The subscript $h$ indicates the spin-sum of arbitrary momentum is computed in the helicity basis which differs from $N(\p)$. Whenever confusions may arise, any functions in the helicity basis will be accompanied by the subscript $h$.

While there is a well-defined procedure to compute $S(\theta,\phi)$ by diagonalizing $\J\cdot\hat{\p}$,  this is not a practical task for large matrices. Instead, it is easier to determine $S\Theta S^{\dag}$ directly. Using eq.~(\ref{eq:Theta}) where $\Theta J_{3}\Theta^{-1}=-J_{3}$, we obtain the following condition
\begin{equation}
\{G,\J\cdot\hat{\p}\}=O \label{eq:anticommute}
\end{equation}
where $G=S\Theta S^{\dag}$ must anti-commute with $\J\cdot\hat{\p}$. However, eq.~(\ref{eq:anticommute}) can only determine $G$ up to a constant. To determine the constant, we use the $j=\frac{1}{2}$ spin-sum as an initial theoretical data and then assume these properties hold for any spin. When $j=\frac{1}{2}$, the self-conjugate and anti-self-conjugate spin-sums, with all the proportionality constants determined are
\begin{equation}
N_{h}^{(j=1/2)}(\p)=\frac{m}{2}\left[I+\mathcal{G}^{(j=1/2)}(\p)\right]\nonumber
\end{equation}
\begin{equation}
M_{h}^{(j=1/2)}(\p)=\frac{m}{2}\left[-I+\mathcal{G}^{(j=1/2)}(\p)\right] \label{eq:spin_half_sums}
\end{equation}
where $\mathcal{G}^{(j=1/2)}(\p)$ is an off-diagonal matrix of the form
\begin{equation}
\mathcal{G}^{(j=1/2)}(\p)=i\left(\begin{matrix}
0 & 0 & 0 & -e^{-i\phi} \\
0 & 0 & e^{i\phi} & 0 \\
0 & -e^{-i\phi} & 0 & 0\\
e^{i\phi} & 0 & 0 & 0 \end{matrix}\right).\label{eq:g_phi}
\end{equation}
We will determine all the proportionality constants of Elko in sec.~\ref{field_operators} such that the resulting spin-sums coincide with eq.~(\ref{eq:spin_half_sums}) when $j=\frac{1}{2}$. As a result of our assumption, the matrix $G$ must be off-diagonal for any spin. An evidence in support of this result is that $\lim_{\theta,\phi\rightarrow 0}G(\theta,\phi)=\Theta$ is an off-diagonal matrix. Therefore, we take $G$ to be
\begin{equation}
G_{\ell m}=g_{\ell}(\theta,\phi)\delta_{\ell,-m}.\label{eq:component_A}
\end{equation}
Substituting eqs.~(\ref{eq:component_A}) and (\ref{eq:r3}) into eq.~(\ref{eq:anticommute}), we obtain the recursive relation
\begin{equation}
g_{-\sigma}(\theta,\phi)=-g_{-\sigma+1}(\theta,\phi)e^{2i\phi}.
\end{equation}
Equating the spin-sums, modulo their respective undetermined proportionality constants in $\mathcal{G}^{(j=1/2)}(\p)$, we get $g_{-1/2}=e^{\phi}$. Extrapolate this to all spin, we obtain
\begin{equation}
g_{-j}=e^{2ij\phi}.\label{eq:g_-j}
\end{equation}
and the following conditions
\begin{equation}
i\beta(j)c^{*2}=1,\hspace{0.5cm}
i\beta(j)d^{*2}=1.
\end{equation}
The Elko norms then become
\begin{subequations}
\begin{equation}
\dual{\xi}(\p,\sigma)\xi(\p,\sigma')=m\delta_{\sigma\sigma'},
\end{equation}
\begin{equation}
\dual{\zeta}(\p,\sigma)\zeta(\p,\sigma')=-m\delta_{\sigma\sigma'}.
\end{equation}
\end{subequations}
Solving the recursive relation using eq.~(\ref{eq:g_-j}), the solution of $g_{\ell}(\phi)$ is given by
\begin{equation}
g_{\ell}=(-1)^{j+\ell}e^{-2i\ell\phi}
\end{equation}
where $\ell=-j,\cdots j$ so that
\begin{equation}
G_{\ell m}=(-1)^{j+\ell}e^{-2i\ell\phi}\delta_{\ell,-m}.\label{eq:A_soln}
\end{equation}
For example, when $j=\frac{1}{2}$, we have
\begin{equation}
G^{(j=1/2)}=\left(\begin{matrix}
0 & -e^{-i\phi} \\
e^{i\phi} & 0 \end{matrix}\right)
\end{equation}
and for $j=1$,
\begin{equation}
G^{(j=1)}=\left(\begin{matrix}
0 & 0 & e^{-2i\phi}\\
0 & -1 & 0 \\
e^{2i\phi} & 0 & 0 \end{matrix}\right).
\end{equation}
Therefore, the self-conjugate and anti-self-conjugate spin-sums simplify to
\begin{subequations}
\begin{equation}
N_{h}(\p)=\sum_{\sigma}\xi_{h}(\p,\sigma)\gdual{\xi}_{h}(\p,\sigma)=\frac{m}{2}
\left[I+\mathcal{G}(\p)\right],
\end{equation}
\begin{equation}
M_{h}(\p)=\sum_{\sigma}\zeta_{h}(\p,\sigma)\gdual{\zeta}_{h}(\p,\sigma)=\frac{m}{2}\left[-I
+\mathcal{G}(\p)\right]
\end{equation}
\end{subequations}
where
\begin{equation}
\mathcal{G}(\p)=\beta(j)(-1)^{2j}\left(\begin{matrix}
O & G \\
G & O \end{matrix}\right)\label{eq:mathcalA}
\end{equation}
for any spin and the proportionality constants are set to
\begin{equation}
|c|^{2}=|d|^{2}=1.
\end{equation}

Now  we return to the problem of determining Elko and its spin-sums in the polarization basis. This can now be solved by noting that the spin-sums in the helicity basis specify a preferred plane with only $\phi$-dependence so the limit $\theta\rightarrow0$ does not affect the spin-sum. In this limit, $\phi(\ep,\sigma)\vert_{\theta\rightarrow0}$ becomes a spinor in the polarization basis in the sense that it is now an eigenspinor of $J_{3}$
\begin{equation}
J_{3}\phi(\ep,\sigma)\vert_{\theta\rightarrow0}=\sigma \phi(\ep,\sigma)\vert_{\theta\rightarrow0}.
\end{equation}
The remaining $\phi$-dependence for $\phi(\ep,\sigma)\vert_{\theta\rightarrow0}$ now appears as non-trivial phases. In the spin-half case, these phases ensure the locality of the fields.  Therefore, the self-conjugate and anti-self-conjugate Elko for any spin in the polarization basis are given by
\begin{subequations}
\begin{equation}
\xi(\0_{\phi},\sigma)=\xi(\ep,\sigma)\vert_{\theta\rightarrow0} ,
\end{equation}
\begin{equation}
\zeta(\0_{\phi},\sigma)=\zeta(\ep,\sigma)\vert_{\theta\rightarrow0}
\end{equation}
\end{subequations}
where subscripts denote the directional-dependence of the rest functions. Similarly the corresponding Elko of arbitrary momentum is denoted as
\begin{subequations}
\begin{equation}
\xi(\p_{\phi},\sigma)=\xi_{h}(\p,\sigma)\vert_{\theta\rightarrow0},
\end{equation}
\begin{equation}
\zeta(\p_{\phi},\sigma)=\zeta_{h}(\p,\sigma)\vert_{\theta\rightarrow0}.
\end{equation}
\end{subequations}
In the helicity basis, Elko of arbitrary momentum satisfy
\begin{subequations}
\begin{equation}
\xi_{h}(\p,\sigma)=\mathcal{S}\kappa(p^{3})\xi(\0,\sigma)=\mathcal{S}\xi(p^{3},\sigma),\label{eq:helicity_xi_p}
\end{equation}
\begin{equation}
\zeta_{h}(\p,\sigma)=\mathcal{S}\kappa(p^{3})\zeta(\0,\sigma)=\mathcal{S}\xi(p^{3},\sigma).\label{eq:helicity_zeta_p}
\end{equation}
\end{subequations}
On the left-hand side we have Elko of arbitrary momentum in the helicity basis while Elko on the right-hand side are defined in the polarization basis according to eq.~(\ref{eq:phi0}). From now onwards, we drop the subscript $h$ as the results obtained for the rest of the paper hold in both helicity and polarization basis.

To conclude this section, the introduction of direction-dependent non-trivial phases ensure the Elko spin-sums remain the same in both helicity and polarization basis. Our results show that the Elko spin-sums for any spin are not Lorentz-covariant and contain a preferred plane. Consequently, the resulting quantum fields violate Lorentz symmetry. Determining the underlying symmetry for such quantum fields is undoubtedly an important task, but it is not one we will undertake in this paper since our starting point is not suitable for this task. 
A potentially more appropriate approach is to start with a symmetry group and derive the particle states and quantum fields following the formalism developed by Wigner and Weinberg~\cite{Wigner:1939cj,Weinberg:1964cn,Weinberg:1964ev}. 

On this front, progress has been made by Ahluwalia and Horvath~\cite{Ahluwalia:2010zn}. They showed that the Elko spin-sums are covariant under VSR transformations proposed by Cohen and Glashow~\cite{Cohen:2006ky} thus suggesting Elko satisfies VSR symmetry. However, to show that the fields for any spin satisfy VSR symmetry, one must derive Elko from first principle. This task has not yet been accomplished. Therefore, we leave this as an open problem for future investigation. For now, in the context of this paper, the next task is to derive the equations of motion for Elko in the momentum space.

\subsection{Higher-spin field equations}


The field equation can be derived by exploiting the following general identity in the $(j,0)\oplus(0,j)$ representation
\begin{equation}
\kappa(p)\Gamma\kappa^{-1}(\p)=\left(\begin{matrix}
O & \exp(2\J\cdot\bv) \\
\exp(-2\J\cdot\bv) & O \end{matrix}\right)\equiv
\frac{(-1)^{2j}}{m^{2j}}\gamma^{\mu_{1}\cdots\mu_{2j}}p_{\mu_{1}}\cdots p_{\mu_{2j}},\label{eq:gamma_2j}
\end{equation}
\begin{equation}
\gamma^{\mu_{1}\cdots\mu_{2j}}\gamma^{\nu_{1}\cdots\nu_{2j}}p_{\mu_{1}}\cdots p_{\mu_{2j}}p_{\nu_{1}}\cdots p_{\nu_{2j}}=
(p^{\mu}p_{\mu})^{2j}=m^{4j}.\label{eq:generalised_dispersion}
\end{equation}
where $\gamma^{\mu_{1}\cdots\mu_{2j}}$ is a generalization of the spin-half $\gamma^{\mu}$ matrices~\cite[app.~A]{Weinberg:1964cn}. These matrices are completely symmetric in all indices and is a rank-$2j$ tensor with respect to the Lorentz transformation of the $(j,0)\oplus(0,j)$ representation.
Here we shall work with $\chi(\p,\sigma)$ leaving the phase $\vartheta$ unfixed for convenience. This is possible because it does not affect the final results. To see how eq.~(\ref{eq:gamma_2j}) helps us, we compute $\Gamma\chi(\0,\sigma)$ and find
\begin{equation}
\Gamma\chi(\0,\sigma)=\pm\vartheta f(j)(-1)^{-j+\sigma} \chi(\0,-\sigma)\label{eq:rest_equation_chi}
\end{equation}
where $f(j)$ is defined as
\begin{equation}
f(j)=\begin{cases}
1 & j=\frac{1}{2},\frac{3}{2},\cdots\\
i & j=1,2,\cdots
\end{cases}
\end{equation}
To derive eq.~(\ref{eq:rest_equation_chi}), it is easier to consider the bosonic and fermionic Elko separately. The $\pm$ sign originates from taking the square root of eq.~(\ref{eq:cd_constants})
\begin{equation}
c^{*}=\pm(-1)^{j+1/2}c,\hspace{0.5cm}
d^{\,*}=\pm(-1)^{j+1/2}d.
\end{equation}
For generality, we keep the freedom of the $\pm$ sign but the choice is inconsequential since the field equation is insensitive to them. Multiply eq.~(\ref{eq:rest_equation_chi}) from the left by $\kappa(\p)$ and using eq.~(\ref{eq:gamma_2j}), we obtain
\begin{subequations}
\begin{equation}
\gamma^{\mu_{1}\cdots\mu_{2j}}p_{\mu_{1}}\cdots p_{\mu_{2j}}\chi(\p,\sigma)=\pm f(j)(-1)^{j+\sigma}\vartheta m^{2j}\chi(\p,-\sigma) \label{eq:chi_eq1}
\end{equation}
\begin{equation}
\gamma^{\mu_{1}\cdots\mu_{2j}}p_{\mu_{1}}\cdots p_{\mu_{2j}}\chi(\p,-\sigma)=\mp f^{-1}(j)(-1)^{-j-\sigma}\vartheta m^{2j}\chi(\p,\sigma).\label{eq:chi_eq2}
\end{equation}
\end{subequations}
The above equations show that Elko for any spin does not satisfy the usual Lorentz-invariant higher-spin equations.\footnote{See~\ref{A} for details.} Instead the operator $\gamma^{\mu_{1}\cdots\mu_{2j}}p_{\mu_{1}}\cdots p_{\mu_{2j}}$ maps $\chi(\p,\sigma)$ to $\chi(\p,-\sigma)$. Substituting eq.~(\ref{eq:chi_eq2}) into eq.~(\ref{eq:chi_eq1}) or vice versa, we obtain
\begin{equation}
\left(\gamma^{\mu_{1}\cdots\mu_{2j}}\gamma^{\nu_{1}\cdots\nu_{2j}}p_{\mu_{1}}\cdots p_{\mu_{2j}}p_{\nu_{1}}\cdots p_{\nu_{2j}}-m^{4j}I\right)\chi(\p,\sigma)=0.
\end{equation}
From eq.~(\ref{eq:generalised_dispersion}), this becomes a higher-spin generalization of the Klein-Gordon equation
\begin{equation}
[(p^{\mu}p_{\mu})^{2j}-m^{4j}]\chi(\p,\sigma)=0
\end{equation}
and consequently
\begin{equation}
[(p^{\mu}p_{\mu})^{2j}-m^{4j}]\xi(\p,\sigma)=[(p^{\mu}p_{\mu})^{2j}-m^{4j}]\zeta(\p,\sigma)=0.\label{eq:Elko_KG}
\end{equation}



\section{Mass-dimension one fields} \label{field_operators}

In this section, we construct quantum fields using Elko in the helicity basis as expansion coefficients. The unusual properties of Elko that we have encountered are naturally inherited by the field operators and will be revealed once we derive the propagators. On the other hand, deriving the Lagrangian density is highly non-trivial and is consequently beyond the scope of this paper. In secs.~\ref{kinematics}  and~\ref{spin_statistics}, it is shown that the naive Klein-Gordon Lagrangian density deduced from eq.~(\ref{eq:Elko_KG})  does not provide an adequate description of the mass dimension one fields. 

We take the field operator with the appropriate normalization to be
\begin{equation}
\Lambda(x)=(2\pi)^{-3/2}\int\frac{d^{3}p}{\sqrt{2mE_{\mathbf{p}}}}\sum_{\sigma}
\left[e^{-ip\cdot x}\xi(\p,\sigma)a(\p,\sigma)+e^{ip\cdot x}\zeta(\p,\sigma)b^{\ddag}(\p,\sigma)\right]
\end{equation}
where the creation and annihilation operators satisfy the canonical algebra
\begin{equation}
[a(\p,\sigma),a^{\ddag}(\p',\sigma')]_{\pm}=[b(\p,\sigma),b^{\ddag}(\p',\sigma')]_{\pm}=\delta_{\sigma\sigma'}\delta^{3}(\p-\p')
\label{eq:creation_annihilation}
\end{equation}
while all other combinations vanish. The particles and anti-particles are created by acting $a^{\ddag}(\p,\sigma)$ and $b^{\ddag}(\p,\sigma)$ on the vacuum $|\,\,\rangle$. The operator $\ddag$ has the same operational meaning as Hermitian conjugation when applied to the particle states but its action on the coefficients is different. We define its operation on Elko as
\begin{subequations}
\begin{equation}
\xi^{\ddag}(\p,\sigma)\equiv\gdual{\xi}(\p,\sigma)\Gamma,
\end{equation}
\begin{equation}
\zeta^{\ddag}(\p,\sigma)\equiv\gdual{\zeta}(\p,\sigma)\Gamma.
\end{equation}
\end{subequations}
Therefore, the adjoint of the quantum field is given by
\begin{equation}
\dual{\Lambda}(x)\equiv(2\pi)^{-3/2}\int\frac{d^{3}p}{\sqrt{2mE_{\mathbf{p}}}}\sum_{\sigma}
\left[e^{ip\cdot x}\gdual{\xi}(\p,\sigma)a^{\ddag}(\p,\sigma)+e^{-ip\cdot x}\gdual{\zeta}(\p,\sigma)b(\p,\sigma)\right].
\label{eq:L_Lad}
\end{equation}

The work of Ahluwalia and Horvath~\cite{Ahluwalia:2010zn} suggested that Elko satisfies VSR symmetry proposed by Cohen and Glashow~\cite{Cohen:2006ky}. Since the VSR groups are subgroups of the Poincar\'{e} group, we expect $\Lambda(x)$ and $\dual{\Lambda}(x)$ to preserve certain properties of Lorentz-invariant field theories. In this spirit, we compute $[\Lambda(\x,t),\dual{\Lambda}(\y,t)]_{\pm}$ to examine their spin-statistics and locality structure. Using eq.~(\ref{eq:creation_annihilation}), we get
\begin{equation}
[\Lambda(\x,t),\dual{\Lambda}(\y,t)]_{\pm}=(2\pi)^{-3}
\int\frac{d^{3}p}{2mE_{\mathbf{p}}}e^{i\mathbf{p\cdot(x-y)}}\left[N(\p)\pm M(-\p)\right].
\end{equation}
Using the fact that $\mathcal{G}(\p)=\pm\mathcal{G}(-\p)$ where the top and bottom sign applies for the bosonic and fermionic spin-sums respectively, for the fermionic fields, we obtain
\begin{equation}
[\Lambda(\x,t),\dual{\Lambda}(\y,t)]_{+}=O,\hspace{0.5cm} j=\frac{1}{2},\frac{3}{2},\cdots.
\end{equation}
As for the bosonic fields, we get
\begin{eqnarray}
[\Lambda(\x,t),\dual{\Lambda}(\y,t)]_{-}&=&(2\pi)^{-3}\int\frac{d^{3}p}{4E_{\mathbf{p}}}e^{i\mathbf{p\cdot(x-y)}}I \nonumber\\
&=&\frac{m}{8\pi^{2}|\x|}K_{1}(m|\x|),\hspace{0.5cm}j=1,2,\cdots.\label{eq:A_integral}
\end{eqnarray}
where $K_{1}(m|\x|)$ is the standard Hankel function. Its absolute value vanishes in the limit $|\x|\rightarrow\infty$ but it is non-vanishing for finite $|\x|$ thus making the bosonic fields non-local.
\subsection{The propagators}\label{kinematics}

We are now ready to derive the propagator. Using the quantum field and its adjoint, the propagator may be obtained by computing $\langle\,\,|T[\Lambda(x)\gdual{\Lambda}(y)]|\,\,\rangle$. Substituting the Elko spin-sums into the generic expression of the propagator given by eq.~(\ref{eq:g_prop}), we obtain the bosonic and fermionic propagators
\begin{subequations}
\begin{equation}
S(y,x)=\frac{i}{2}\int\frac{d^{4}q}{(2\pi)^{4}}e^{-iq\cdot(x-y)}
\left(\frac{q^{0}}{E_{\mathbf{q}}}\right)
\left[\frac{I+(E_{\mathbf{q}}/q^{0})\mathcal{G}(\q)}{q^{\mu}q_{\mu}-m^{2}+i\epsilon}\right],\,j=1,2,\cdots \label{eq:bosonic_propagator}
\end{equation}
\begin{equation}
S(y,x)=
\frac{i}{2}\int\frac{d^{4}q}{(2\pi)^{4}}e^{-iq\cdot(x-y)}\left[\frac{I+\mathcal{G}(\q)}{q^{\mu}q_{\mu}-m^{2}+i\epsilon}\right], \, j=\frac{1}{2},\frac{3}{2},\cdots. \label{eq:fermionic_propagator}
\end{equation}
\end{subequations}
Equations~(\ref{eq:bosonic_propagator}) and (\ref{eq:fermionic_propagator}) are the main results of this paper. The integrals over $\mathcal{G}(\q)$ for the bosonic and fermionic propagators are generally non-zero unless $\mathbf{x-y}$ is aligned along the preferred 3-axis. A simple dimensional analysis shows that the fields for any spin all have mass-dimension one. This result is in stark contrast with the well-known result that the massive Lorentz-invariant fields of the $(j,0)\oplus(0,j)$ representation are of mass-dimension $j+1$. But in hindsight, since Lorentz symmetry is violated, the general properties of Lorentz-invariant field theories may no longer be applicable.

\subsection{Canonical variables}\label{spin_statistics}

The particles and anti-particles we have introduced are by construction, physically distinct. As a result, the corresponding Lagrangian density has two independent variables $\Lambda(x)$ and $\dual{\Lambda}(x)$ and are accompanied by their respective conjugate momenta $\Pi(x)$ and $\dual{\Pi}(x)$.

We define the conjugate momenta to be
\begin{equation}
\Pi(x)\equiv\frac{\partial\dual{\Lambda}}{\partial t}(x),\hspace{0.5cm}
\dual{\Pi}(x)\equiv(-1)^{2j}\frac{\partial\Lambda}{\partial t}(x).
\label{eq:conjugate_momenta}
\end{equation}
Modulo the $(-1)^{2j}$ factor for $\dual{\Pi}(x)$, the conjugate momenta can be obtained from the 
Klein-Gordon Lagrangian density. In sec.~\ref{free_hamiltonian}, we show that the factor $(-1)^{2j}$ is required in order to obtain the positive-definite Hamiltonians. The canonical variables are taken to be
\begin{subequations}
\begin{equation}
Q_{1}(x)=\Lambda(x),\hspace{0.5cm}
Q_{2}(x)=\dual{\Lambda}(x),\label{eq:canonical1}
\end{equation} 
\begin{equation}
P_{1}(x)=\Pi(x),\hspace{0.5cm}
P_{2}(x)=\dual{\Pi}(x). \label{eq:canonical2}
\end{equation}
\end{subequations}
A straightforward computation gives us the following bosonic commutators
\begin{subequations}
\begin{equation}
[\Lambda(\x,t),\Pi(\y,t)]_{-}=\frac{i}{2}\int\frac{d^{3}p}{(2\pi)^{3}}
e^{i\mathbf{p\cdot(x-y)}}\mathcal{G}(\p),\label{eq:anti_commutator1}
\end{equation}
\begin{equation}
[\dual{\Lambda}(\x,t),\dual{\Pi}(\y,t)]=-\frac{i}{2}\int\frac{d^{3}p}{(2\pi)^{3}}e^{-i\mathbf{p\cdot(x-y)}}\mathcal{G}^{T}(\p),\hspace{0.5cm}j=1,2,\cdots.\label{eq:anti_commutator2}
\end{equation}
\end{subequations}
The fermionic anti-commutators are
\begin{subequations}
\begin{equation}
[\Lambda(\x,t),\Pi(\y,t)]_{+}=\frac{i}{2}\int\frac{d^{3}p}{(2\pi)^{3}}e^{i\mathbf{p\cdot(x-y)}}[I+\mathcal{G}(\p)], \label{eq:anti_commutator3}
\end{equation}
\begin{equation}
[\dual{\Lambda}(\x,t),\dual{\Pi}(\y,t)]_{+}=\frac{i}{2}\int\frac{d^{3}p}{(2\pi)^{3}}
e^{-i\mathbf{p\cdot(x-y)}}[I+\mathcal{G}^{T}(\p)]\hspace{0.5cm}j=\frac{1}{2},\frac{3}{2},\cdots.\label{eq:anti_commutator4}
\end{equation}
\end{subequations}

The above analysis has revealed the unsatisfactory aspects of the Klein-Gordon Lagrangian density. The problem resides in the fact that the integrals of $\mathcal{G}(\p)$ are non-vanishing unless $\x-\y$ is aligned to the preferred 3-axis. On this front, a new Lagrangian density which resolves these problems has been proposed for the case of $j=\frac{1}{2}$.~\cite{Lee:2014opa}. This Lagrangian density can in principle, be generalized to higher spin fermionic fields. Unfortunately, a similar extension to higher-spin bosonic fields is not possible. For now, we shall leave the task of deriving the correct Lagrangian density to a future publication.

\subsection{The Hamiltonians}\label{free_hamiltonian}
Despite the problems associated with the Klein-Gordon Lagrangian density, we show that the canonical variables defined in eqs.~(\ref{eq:canonical1}-\ref{eq:canonical2}) do give us positive-definite Hamiltonians for any spin. Consequently, the Hamiltonians obtained from any new Lagrangian densities must be identical to the Klein-Gordon Hamiltonian. 

The Hamiltonians are evaluated using the Legendre transformation
\begin{equation}
H_{0}
=\int d^{3}x \left[
(-1)^{2j}\frac{\partial\Lambda}{\partial t}\frac{\partial\dual{\Lambda}}{\partial t} -\partial_{i}\dual{\Lambda}\partial^{i}\Lambda+m^{2}\dual{\Lambda}\Lambda\right].
\end{equation}
Substituting $\Lambda(x)$ and $\dual{\Lambda}(x)$ into $H_{0}$, we obtain
\begin{eqnarray}
H_{0}&=&\int\frac{d^{3}p}{m}E_{\mathbf{p}}\sum_{\sigma}\Big[\dual{\xi}(\p,\sigma)\xi(\p,\sigma)[a^{\ddag}(\p,\sigma)a(\p,\sigma)+\delta(\0)]\nonumber\\
&&\hspace{3cm}+\dual{\zeta}(\p,\sigma)\zeta(\p,\sigma)[(-1)^{2j}b^{\ddag}(\p,\sigma)b(\p,\sigma)+\delta(\0)]\Big].
\end{eqnarray}
It is important to remark that if the $(-1)^{2j}$ factor is absent from $\dual{\Pi}(x)$, the cross-terms proportional to $a(\p,\sigma)b(-\p,\sigma')$ and $a^{\ddag}(\p,\sigma)b^{\ddag}(-\p,\sigma')$ would be non-vanishing. Therefore, the introduction of $(-1)^{2j}$ to $\dual{\Pi}(x)$ is crucial in ensuring that the  fermionic Hamiltonian is diagonal. Also, we see that the norms of the self-conjugate and anti-self-conjugate Elko must be identical within their respective sectors so that they contribute equally to the Hamiltonian.

Using eq.~(\ref{eq:cd_constants}) and the Elko norms given in eqs.~(\ref{eq:xi_norm_final}-\ref{eq:zeta_norm_final}), for the fermionic Hamiltonian, the divergent terms proportional to $\delta^{3}(\0)$ cancel giving us
\begin{equation}
H_{0}=i\beta(j)c^{*2}\int d^{3}p\sum_{\sigma}E_{\mathbf{p}}\left[a^{\ddag}(\p,\sigma)a(\p,\sigma)
+b^{\ddag}(\p,\sigma)b(\p,\sigma)\right],\hspace{0.5cm}j=\frac{1}{2},\frac{3}{2},\cdots.
\end{equation}
Therefore, the fermionic field for Elko has vanishing vacuum energy. Setting the proportionality constants of the fermionic field and its adjoint to be
\begin{equation}
\beta(j)=-i,\hspace{0.5cm} c=d=1,\hspace{0.5cm}j=\frac{1}{2},\frac{3}{2},\cdots
\end{equation}
we obtain the desired fermionic Hamiltonian with no vacuum energy
\begin{equation}
H_{0}=\int d^{3}p\sum_{\sigma}E_{\mathbf{p}}\left[a^{\ddag}(\p,\sigma)a(\p,\sigma)
+b^{\ddag}(\p,\sigma)b(\p,\sigma)\right],\hspace{0.5cm}j=\frac{1}{2},\frac{3}{2},\cdots.
\end{equation}
Performing a similar computation for the bosonic field, in order to obtain a physical Hamiltonian, we set the proportionality constants to
\begin{equation}
\beta(j)=1,\hspace{0.5cm} c=d^{*}=e^{i\pi/4},\hspace{0.5cm} j=1,2,\cdots
\end{equation}
which gives us
\begin{equation}
H_{0}=\int d^{3}p\sum_{\sigma}E_{\mathbf{p}}\left[a^{\ddag}(\p,\sigma)a(\p,\sigma)
+b^{\ddag}(\p,\sigma)b(\p,\sigma)+2\delta^{3}(\0)\right],\hspace{0.5cm}j=1,2,\cdots.
\end{equation}
Unlike the fermions, the vacuum energy is positive and non-vanishing, but since only the shift in vacuum energy is measurable, the divergent term can be ignored. It is helpful to remind ourselves that it is unusual for the fermionic field to have vanishing vacuum energy. In Lorentz-invariant field theories, the Hamiltonians of bosons and fermions have positive and negative divergent vacuum energies respectively.

\subsection{A decomposition of $\gamma^{\mu_{1}\cdots\mu_{2j}}p_{\mu_{1}}\cdots p_{\mu_{2j}}$}

We have now determined all the proportionality constants so it is only appropriate that the important results obtained in this paper are presented in their final forms. It is certainly possible and without any loss of generality, to choose these constants at the beginning of our construction. However, this would have left the reader wondering how they are determined and whether the choices are arbitrary. Our derivation shows that, apart from trivial global phases, there are no additional freedoms. 

Utilizing the results we have obtained so far, we are able to generalize the results obtained by Ahluwalia~\cite{Ahluwalia:2013uxa} and Speran\c{c}a~\cite{Speranca:2013hqa} by decomposing $\gamma^{\mu_{1}\cdots\mu_{2j}}p_{\mu_{1}}\cdots p_{\mu_{2j}}$. 
Starting with the Elko spin-sums, we have
\begin{equation}
\sum_{\sigma}\chi(\p,\sigma)\dual{\chi}(\p,\sigma)=\frac{m}{2}[-i\vartheta I+\mathcal{G}(\p)]
\end{equation}
where the self-conjugate and anti-self-conjugate spin-sums can be obtained by taking $\vartheta=i$ and $\vartheta=-i$ respectively. There is in fact an elegant way to relate the spin-sums to $\gamma^{\mu_{1}\cdots\mu_{2j}}p_{\mu_{1}\cdots\mu_{2j}}$. For this purpose, we go back to eqs.~(\ref{eq:chi_eq1}) which allows us to derive the following spin-sum for all spin
\begin{equation}
\sum_{\sigma}\chi(\p,\sigma)\bar{\chi}(\p,\sigma)=\frac{1}{2m^{2j-1}}
[I-i\vartheta\mathcal{G}(\p)]\gamma^{\mu_{1}\cdots\mu_{2j}}p_{\mu_{1}}\cdots p_{\mu_{2j}}.
\end{equation}

Let us now introduce the operator~\cite{Speranca:2013hqa,Ahluwalia:2013uxa}
\begin{eqnarray}
\Xi(\p)&=&\frac{1}{m}\sum_{\sigma}\left[\xi(\p,\sigma)\bar{\xi}(\p,\sigma)-\zeta(\p,\sigma)\bar{\zeta}(\p,\sigma)\right]\nonumber\\
&=&\frac{1}{m^{2j}}\mathcal{G}(\p)\gamma^{\mu_{1}\cdots\mu_{2j}}p_{\mu_{1}}\cdots p_{\mu_{2j}}.
\end{eqnarray}
We can now derive some interesting identities. Consider
\begin{equation}
\Xi^{2}(\p)=\frac{1}{m^{4j}}\mathcal{G}(\p)\gamma^{\mu_{1}\cdots\mu_{2j}}p_{\mu_{1}}\cdots p_{\mu_{2j}}\mathcal{G}(\p)
\gamma^{\nu_{1}\cdots\nu_{2j}}p_{\nu_{1}}\cdots p_{\nu_{2j}}
\end{equation}
Using eqs.~(\ref{eq:anticommute}) and (\ref{eq:mathcalA}), it is straightforward to show that 
\begin{equation}
[\mathcal{G}(\p),\gamma^{\mu_{1}\cdots\mu_{2j}}p_{\mu_{1}}\cdots p_{\mu_{2j}}]=O.\label{eq:A_gamma}
\end{equation}
It follows that
\begin{equation}
\Xi^{2}(\p)=\mathcal{G}^{2}(\p)=I
\end{equation}
and
\begin{equation}
[\Xi(\p),\mathcal{G}(\p)]=[\Xi(\p),\gamma^{\mu_{1}\cdots\mu_{2j}}p_{\mu_{1}}\cdots p_{\mu_{2j}}]=O.\label{eq:Xi_A_Gamma}
\end{equation}
Equations (\ref{eq:A_gamma}) and (\ref{eq:Xi_A_Gamma}) were initially derived by Ahluwalia for $j=\frac{1}{2}$~\cite{Ahluwalia:2013uxa}. Our result shows that these identities are valid for any spin. 

An interesting way to look these identities is that $\gamma^{\mu_{1}\cdots\mu_{2j}}p_{\mu_{1}}\cdots p_{\mu_{2j}}$ can be decomposed into a product of two commuting matrices
\begin{equation}
\gamma^{\mu_{1}\cdots\mu_{2j}}p_{\mu_{1}}\cdots p_{\mu_{2j}}=m^{2j}\mathcal{G}(\p)\Xi(\p).
\end{equation}
The matrices $\Xi(\p)$ and $\mathcal{G}(\p)$ are non-covariant but the violating terms cancel leaving their product Lorentz-covariant. Our observation suggest that there are potentially interesting mathematics behind the decomposition. 
For now, we leave their transformation properties as a future subject of research.

\section{Conclusions}

In Refs.~\citen{Ahluwalia:2004ab,Ahluwalia:2004sz,Ahluwalia:2008xi,Ahluwalia:2009rh}, it was shown that contrary to conventional wisdom, there exists fermionic fields of mass-dimension one that satisfy the Klein-Gordon but not the Dirac equation. These fields constructed using spin-half Elko as expansion coefficients are Lorentz-violating and have renormalizable self-interactions. The results obtained in this paper are a direct generalization of these works. The properties of spin-half Elko are naturally inherited by their higher-spin generalizations. 

It should be noted that our work may not be of pure academic interest. The mass dimension one fermions of spin-half are potential dark matter candidates due to renormalizable self-interactions and limited electromagnetic interactions. Since these properties remain unchanged for the higher-spin fields, it follows that they are also likely dark matter candidates. If dark matter exists, they are unlikely to be comprised entirely of one single specie. Our construct thus offers a possible solution since it contains both bosonic and fermionic particles. 

While there are stringent limits on Lorentz symmetry violation, they cannot rule out our construct as these experiments were conducted using only the SM particles. There are no direct evidence suggesting that dark matter must satisfy Lorentz symmetry~\cite{Blas:2012vn}. Towards determining the symmetries of the mass-dimension one fields, the works of Ahluwalia~\cite{Ahluwalia:2013uxa}, Speran\c{c}a~\cite{Speranca:2013hqa} and our generalization of their results on the decomposition of $\gamma^{\mu_{1}\cdots\mu_{2j}}p_{\mu_{1}}\cdots p_{\mu_{2j}}$, have provided additional insights on Lorentz violation. We believe further investigation in this direction will not only give us a deeper understanding of Elko but also elucidate on a possible connection between mass-dimension one fields and Lorentz-invariant fields.


%
%
%
%
\section*{Acknowledgement}

This research is supported by CNPq grant 313285/2013-6. I would like to thank D. V. Ahluwalia for numerous discussions and for communicating the result on the integrals of $\mathcal{G}(\p)$. I am grateful to the hospitality of the Department of Physics and Astronomy at the University of Canterbury where part of this work was completed.

\appendix

\section{Lorentz-invariant fields for any spin} \label{A}

Let $\Psi(x)$ be a Lorentz-invariant massive quantum field of the $(j,0)\oplus(0,j)$ representation
\begin{equation}
\Psi(x)=(2\pi)^{-3/2}\int\frac{d^{3}p}{\sqrt{2E_{\mathbf{p}}}}\sum_{\sigma}
\left[e^{-ip\cdot x}u(\p,\sigma)a(\p,\sigma)+e^{ip\cdot x}v(\p,\sigma)b^{\dag}(\p,\sigma)\right]
\end{equation}
The expansion coefficients at rest determined by Lorentz symmetry with the appropriate normalization are given by
\begin{equation}
u(\0,j)=m^{j}\left(\begin{matrix}
1 \\
0 \\
\vdots \\
0 \\
1 \\
0 \\
\vdots \\
0
\end{matrix}\right),\,
u(\0,j-1)=m^{j}\left(\begin{matrix}
0 \\
1 \\
\vdots \\
0 \\
0 \\
1 \\
\vdots \\
0
\end{matrix}\right),\,\cdots\,,
u(\0,-j)=m^{j}\left(\begin{matrix}
0 \\
\vdots\\
0 \\
1 \\
0 \\
\vdots\\
0 \\
1 \\
\end{matrix}\right)
\end{equation}
\begin{equation}
v(\0,j)=m^{j}\left(\begin{matrix}
0 \\
\vdots\\
0\\
(-1)^{2j+1}\\
0 \\
\vdots\\
0\\
-1
\end{matrix}\right)\,,
v(\0,j-1)=m^{j}\left(\begin{matrix}
0 \\
\vdots\\
(-1)^{2j}\\
0 \\
0 \\
\vdots\\
1\\
0 \end{matrix}\right),\,\cdots ,
v(\0,-j)=m^{j}\left(\begin{matrix}
-1\\
0\\
\vdots\\
0\\
(-1)^{2j+1}\\
0\\
\vdots\\
0\end{matrix}\right).
\end{equation}
The expansion coefficients of arbitrary momentum are obtained by
\begin{subequations}
\begin{equation}
u(\p,\sigma)=\kappa(\p)u(\0,\sigma),
\end{equation}
\begin{equation}
v(\p,\sigma)=\kappa(\p)v(\0,\sigma)
\end{equation}
\end{subequations}
and they satisfy the following field equations
\begin{subequations}
\begin{equation}
(\gamma^{\mu_{1}\cdots\mu_{2j}}p_{\mu_{1}}\cdots p_{\mu_{2j}}-m^{2j})u(\p,\sigma)=0,
\end{equation}
\begin{equation}
[\gamma^{\mu_{1}\cdots\mu_{2j}}p_{\mu_{1}}\cdots p_{\mu_{2j}}-(-1)^{2j}m^{2j}]v(\p,\sigma)=0.
\end{equation}
\end{subequations}
The explicit expression of $\gamma^{\mu_{1}\cdots\mu_{2j}}$ up to spin-two can be found in Ref.~\citen{Ahluwalia:1991gs}.
Consequently, the field equation for $\Psi(x)$ is
\begin{equation}
\left[(i)^{2j}\gamma^{\mu_{1}\cdots\mu_{2j}}\partial_{\mu_{1}}\cdots\partial_{\mu_{2j}}-m^{2j}\right]\Psi(x)=0.
\end{equation}

Our main purpose here is to derive the charge-conjugation matrix $\mathcal{C}$ needed to construct Elko for any spin. To achieve this, we start with the standard definition of charge-conjugation operator $\mathsf{C}$ acting on the creation and annihilation operators
\begin{equation}
\mathsf{C}a(\p,\sigma)\mathsf{C}^{-1}=\varsigma^{*} b(\p,\sigma),\hspace{0.5cm}
\mathsf{C}b^{\dag}(\p,\sigma)\mathsf{C}^{-1}=\bar{\varsigma} a^{\dag}(\p,\sigma)
\end{equation}
where $\varsigma$ and $\bar{\varsigma}$ are the charge-conjugation phases. Therefore, the field $\Psi(x)$ transforms as
\begin{equation}
\mathsf{C}\Psi(x)\mathsf{C}^{-1}=(2\pi)^{-3/2}\int\frac{d^{3}p}{\sqrt{2E_{\mathbf{p}}}}
\sum_{\sigma}\left[e^{-ip\cdot x}\varsigma^{*}u(\p,\sigma)b(\p,\sigma)
+e^{ip\cdot x}\bar{\varsigma}v(\p,\sigma)a^{\dag}(\p,\sigma)\right].
\end{equation}
The expansion coefficients satisfy the following identities
\begin{equation}
\mathcal{C}u(\p,\sigma)=iv(\p,\sigma),\hspace{0.5cm}
\mathcal{C}v(\p,\sigma)=iu(\p,\sigma)\label{eq:Cuv}
\end{equation}
where charge-conjugation matrix is defined as
\begin{equation}
\mathcal{C}=\left(\begin{matrix}
O & -i\Theta^{-1} \\
-i\Theta & O 
\end{matrix}\right)\mathcal{K}
\end{equation}
with $\mathcal{K}$ being the complex conjugation operator acting to the right. Setting $\varsigma^{*}=\bar{\varsigma}$ and using eq.~(\ref{eq:Cuv}), we obtain
\begin{equation}
\mathsf{C}\Psi(x)\mathsf{C}^{-1}=-i\varsigma^{*}\mathcal{C}\Psi(x).
\end{equation}

\section{Spin-half Elko}\label{B}

We define a four-component spinor $\chi(\p,\sigma)$ as
\begin{equation}
\chi(\p,\sigma)=
\left(\begin{matrix}
\vartheta\Theta\phi^{*}(\p,\sigma)\\
\phi(\p,\sigma)\end{matrix}\right)
\end{equation}
where $\Theta=-i\sigma_{2}$. The spinor $\chi(\p,\sigma)$ becomes the eigenspinors of the charge-conjugation operator $\mathcal{C}$ (and hence Elko) with the following choice of phases
\begin{equation}
\mathcal{C}\chi(\p,\sigma)\vert_{\vartheta=\pm i}=\pm\chi(\p,\sigma)\vert_{\vartheta=\pm i}
\end{equation}
thus giving us four Elko spinors.

In the helicity basis, we take the left-handed Weyl spinors at rest to be
\begin{equation}
\phi(\e,{\textstyle{\frac{1}{2}}})=\sqrt{\frac{m}{2}}
\left(\begin{matrix}
\cos(\theta/2)e^{-i\phi/2}\\
\sin(\theta/2)e^{i\phi/2} \end{matrix}\right),
\end{equation}
\begin{equation}
\phi(\e,-{\textstyle{\frac{1}{2}}})=\sqrt{\frac{m}{2}}
\left(\begin{matrix}
-\sin(\theta/2)e^{-i\phi/2} \\
\cos(\theta/2)e^{i\phi/2}\end{matrix}\right).
\end{equation}
The Elko spinors at rest, are divided into the self-conjugate spinors ($\vartheta=i$) and the anti-self-conjugate spinors ($\vartheta=-i$)
\begin{equation}
\xi(\e,\textstyle{\frac{1}{2}})=+\chi(\e,\textstyle{\frac{1}{2}})\vert_{\vartheta=+i},
\end{equation}
\begin{equation}
\xi(\e,\textstyle{-\frac{1}{2}})=+\chi(\e,\textstyle{-\frac{1}{2}})\vert_{\vartheta=+i},
\end{equation}
\begin{equation}
\zeta(\e,\textstyle{\frac{1}{2}})=+\chi(\e,-\textstyle{\frac{1}{2}})\vert_{\vartheta=-i},
\end{equation}
\begin{equation}
\zeta(\e,-\textstyle{\frac{1}{2}})=-\chi(\e,\textstyle{\frac{1}{2}})\vert_{\vartheta=-i}.
\end{equation}
Our labels for the spinors are different to Ref.~\citen{Ahluwalia:2009rh}. For the self-conjugate spinor, $\sigma=\frac{1}{2},-\frac{1}{2}$ only denotes the eigenvalue of its left-handed component $\phi(\e,\sigma)$ with respect to $\frac{1}{2}\s\cdot\hat{\p}$. Whereas in Ref.~\citen{Ahluwalia:2009rh}, $\sigma=\pm\frac{1}{2},\mp\frac{1}{2}$. The top and bottom sign denote the eigenvalues of both the right and left-handed component of the self-conjugate spinors respectively.

In the polarization basis, the Elko spinors takes symbolic form~\cite{Ahluwalia:2009rh}
\begin{equation}
\xi(\e,\textstyle{\frac{1}{2}})=\left(\begin{matrix}
i \Downarrow\\
\Uparrow \end{matrix}\right),\hspace{0.5cm}
\xi(\e,-\textstyle{\frac{1}{2}})=\left(\begin{matrix}
-i\Uparrow \\
\Downarrow \end{matrix}\right)\label{eq:elko_polarization1}
\end{equation}
\begin{equation}
\zeta(\e,\textstyle{\frac{1}{2}})=\left(\begin{matrix}
i\Uparrow\\
\Downarrow\end{matrix}\right),\hspace{0.5cm}
\zeta(\e,-\textstyle{\frac{1}{2}})=-\left(\begin{matrix}
-i\Downarrow\\
\Uparrow
\end{matrix}\right)\label{eq:elko_polarization2}
\end{equation}
where
\begin{equation}
\Uparrow=\sqrt{\frac{m}{2}}\left(\begin{matrix}
e^{-i\phi/2} \\
0 \end{matrix}\right),\hspace{0.5cm}
\Downarrow=\sqrt{\frac{m}{2}}\left(\begin{matrix}
0 \\
e^{i\phi/2} \end{matrix}\right).
\end{equation}
They are obtained by taking the $\theta\rightarrow0$ limit of Elko in the helicity basis. The arrows $\Uparrow$ and $\Downarrow$ represent eigenspinors of $J_{3}=\frac{\sigma_{3}}{2}$ with positive and negative eigenvalues respectively.

\section{Propagator}\label{propagator}

We compute the most general propagator. Let operator $S(y,x)$ where $x=(t,\x)$ and $y=(t',\y)$ be the propagator for a general field $\psi(x)$
\begin{equation}
\psi(x)=(2\pi)^{-3/2}\int\frac{d^{3}p}{\sqrt{2E_{\mathbf{p}}}}\sum_{\sigma}\left[e^{-ip\cdot x}u(\p,\sigma)a(\p,\sigma)+e^{ip\cdot x}v(\p,\sigma)b^{\dag}(\p,\sigma)\right]
\end{equation}
and its adjoint $\gdual{\psi}(x)$
\begin{equation}
\gdual{\psi}(x)=(2\pi)^{-3/2}\int\frac{d^{3}p}{\sqrt{2E_{\mathbf{p}}}}\sum_{\sigma}\left[e^{ip\cdot x}\dual{u}(\p,\sigma)a^{\dag}(\p,\sigma)+e^{-ip\cdot x}\dual{v}(\p,\sigma)b(\p,\sigma)\right].
\end{equation}
The propagator $S(y,x)$ is defined as
\begin{eqnarray}
S(y,x)&=&\langle\,\,|T[\psi(x)\gdual{\psi}(y)]|\,\,\rangle \nonumber \\
&=&\theta(t-t')\langle\,\,|\psi(x)\gdual{\psi}(y)|\,\,\rangle\pm\theta(t'-t)\langle\,\,|\gdual{\psi}(y)\psi(x)|\,\,\rangle\label{eq:syx}
\end{eqnarray}
where $T$ denotes the time-ordered product and $\theta(t)$ is the step function
\begin{equation}
\theta(t)=\begin{cases}
1 & t\geq 0 \\
0 & t<0
\end{cases}.
\end{equation}
The top and bottom signs in eq.~(\ref{eq:syx}) are for the bosonic and fermionic fields respectively. Computing the vacuum expectation values, the non-vanishing matrix elements are
\begin{equation}
\langle\,\,|\psi(x)\gdual{\psi}(y)|\,\,\rangle=(2\pi)^{-3}\int\frac{d^{3}p}{2E_{\mathbf{p}}}e^{-ip\cdot(y-x)}N(\p),
\end{equation}
\begin{equation}
\langle\,\,|\gdual{\psi}(y)\psi(x)|\,\,\rangle=(2\pi)^{-3}\int\frac{d^{3}p}{2E_{\mathbf{p}}}e^{ip\cdot(y-x)}M(\p)
\end{equation}
where
\begin{equation}
N(\p)=\sum_{\sigma}u(\p,\sigma)\dual{u}(\p,\sigma),
\end{equation}
\begin{equation}
M(\p)=\sum_{\sigma}v(\p,\sigma)\dual{v}(\p,\sigma).
\end{equation}
Using the integral representation of the step function
\begin{equation}
\theta(t)=\lim_{\epsilon\rightarrow0^{+}}\int\frac{d\omega}{2\pi i}
\frac{e^{i\omega t}}{\omega-i\epsilon}
\end{equation}
after some algebraic manipulations and performing the change of variables, $\omega=E_{\mathbf{p}}-q^{0}$ and $\p=\mathbf{q}$,
we obtain
\begin{equation}
S(y,x)=i\int\frac{d^{4}q}{(2\pi)^{4}}\frac{e^{-iq\cdot(y-x)}}{(2\sqrt{{|\mathbf{q}|^{2}+m^{2}}})}
\left[\frac{\sqrt{{|\mathbf{q}|^{2}+m^{2}}}[N(\mathbf{q})\pm M(-\mathbf{q})]+q^{0}[N(\mathbf{q})\mp M(-\mathbf{q})]}{q^{\mu}q_{\mu}-m^{2}+i\epsilon}\right].\label{eq:g_prop}
\end{equation}


\end{document}